\documentclass[pre,twocolumn]{revtex4}
\usepackage{graphicx}
\usepackage{amsopn,color}
\usepackage{amsfonts}
\usepackage{amsmath}
\usepackage{latexsym}
\usepackage{amsmath,bm,amssymb}
\usepackage{tikz}
\usetikzlibrary{decorations.pathmorphing}
\usetikzlibrary{patterns,arrows}
\usetikzlibrary{decorations.markings}
\usepackage{pgfplots}
\usepackage{tkz-fct}

\begin{document}

\title{Correlations of fluctuations of two-dimensional flow forced by a random force
\\ on top of a shear flow}

\author{I.V.Kolokolov$^1 \, ^2$, V.V.Lebedev$^1\, ^2$}

\affiliation{$^1$ Landau Institute for Theoretical Physics, RAS, \\
142432, Chernogolovka, Moscow region, Russia; \\
$^2$ National Research  University Higher School of Economics, \\
101000, Myasnitskaya ul. 20, Moscow, Russia.}

\date{\today}

\begin{abstract}

We examine fluctuations of vorticity excited by an external random force in two-dimensional fluid in the presence of a strong external shear flow. The problem is motivated by the analysis of big coherent vortices appearing as a consequence of the inverse energy cascade in a finite box at large Reynolds numbers. We develop the perturbation theory for calculating nonlinear corrections to correlation functions of the flow fluctuations assuming that the external force is short correlated in time. We analyze corrections to the pair correlation function of vorticity and some moments. The analysis enables one to establish validity of the perturbation theory for laboratory experiments and numerical simulations.

\end{abstract}

\maketitle

\section{Introduction}
\label{sec:intro}

Two-dimensional turbulence is high-Reynolds hydrodynamic flow in fluid films on scales larger than the film thickness \cite{boffetta2012two}. From the practical point of view, the most interesting such fluid ``film'' is atmosphere. Of course, atmosphere is very complex object and its detailed description is extremely difficult problem. However, some general features of two-dimensional turbulence could be useful for understanding atmospheric phenomena. Note in this respect the trend of the production of large-scale eddies from small-scale ones thanks to the non-linear hydrodynamic interaction in two-dimensional fluids \cite{67Kra,68Lei,69Bat}. In presence of the external forcing (pumping) the trend leads to formation of the inverse energy cascade at scales larger than the pumping length \cite{KraM}.

In a finite box, the transfer of the energy to large scales leads to formation of big eddies with the diameter of the order of the size of the box \cite{Sommeria86,MCH04}. At some conditions big coherent vortices are formed having large life time. Such coherent vortices were observed both in laboratory experiments \cite{ShaF,FilLevchOrlov17} and in numerical simulations \cite{chertkov2007dynamics}. In the work \cite{laurie2014universal} the flat profile of the mean velocity of the coherent vortex was observed in numerical simulations and some arguments explaining the profile were presented. In the works \cite{KL15,kolokolov2016structure,kolokolov2016velocity,frishman17,kolokolov2017static} the quasilinear regime of the flow fluctuations inside the coherent vortex was utilized, the flat velocity profile was derived and some velocity correlation functions were discussed. In the work \cite{KL20} the criterion of the formation of the coherent vortex was proposed, based on the approach. The prediction is qualitatively confirmed in direct numerical simulations \cite{DFKL22}.

The flow fluctuations inside the coherent vortex are excited on top of its mean flow that locally can be approximated by the shear flow. This is the motivation of the present work where we consider the turbulent two-dimensional flow in an unbound fluid film in the presence of a strong shear flow. In our theoretical scheme, we assume that the external force exciting fluctuations of the flow on top of the static shear flow is short correlated in time. We believe that the model reflects general properties of the flow fluctuations excited by an external random force. However, the case where the correlation time of the exciting force is relatively large (greater than the inverse shear rate) needs a special attention. It is a subject of future investigations.

The mean shear flow breaks space homogeneity. The fact makes calculation of the correlation functions of the flow fluctuations essentially more complicated than in the homogeneous case. However, one can formulate a consistent calculation scheme based on the perturbation series. The series is based on the representation of the correlation function via functional integrals can be constructed like in the quantum field theory. The scheme gives the terms of the perturbation series, that can be represented by corresponding Feynman diagrams. We analyse first terms of the perturbation series and formulate some its general properties. Namely, we calculate fluctuation corrections to the pair correlation function of the vorticity, its second moment and the Reynolds stress tensor. The calculations enable one to establish the validity criterion of the perturbation theory for the case.

We believe that the results obtained for the model can be extended to explain properties of the coherent vortices. Its mean velocity is a differential rotation. At analyzing the fluctuation effects, the mean flow can be approximated as the shear flow for distances from the vortex center much larger than the pumping correlation length. From the other hand, the local shear rate of the mean vortex flow diminishes as the distance increases. It restricts the applicability of our approach from above. Thus, we are aimed to explain the properties of the coherent vortex in some region of the distances from the vortex center that is the most interesting, since just the region determines the features of the coherent vortex.

\section{General relations}
\label{sec:general}

We consider the model of an unbounded two-dimensional fluid where some random flow is excited. We assume that the random flow exists on top of the average static shear flow. We choose the reference system with the first axis directed along the velocity of the average shear flow. Then the velocity of the shear flow has the only first component $V_1=\varSigma x_2$ where $\varSigma$ is the shear rate and $x_2$ is the second coordinate. We are interested in statistical properties of the flow fluctuations and examine effects related to their nonlinear interaction. The fluctuations are described by (two-dimensional) random velocity $\bm v$. The flow is assumed to be incompressible: $\nabla \bm v=0$.

Having in mind thin fluid films, we introduce two dissipative mechanisms: bottom friction and viscosity. Then, to support the shear flow, a regular external static force should be applied to the fluid. Besides, we include into the external force applied to the fluid some random component $\bm f$ exciting the flow fluctuations. The force $\bm f$ is assumed to be a random function of time and coordinates with zero average and possessing statistical properties homogeneous in space and time. We analyze the statistically stationary state of the fluid. The state is homogeneous along the first axis, however, the homogeneity along the second axis is broken thanks to the presence of the shear flow.

The equation controlling the flow velocity in the fluid is two-dimensional Navier-Stokes equation with an additional term describing bottom friction. Extracting the equation for the random velocity $\bm v$ on top of the shear flow we find
\begin{eqnarray}
\partial_t \bm v +\varSigma x_2 \partial_1 \bm v
+\varSigma \bm n v_2 +(\bm v \nabla) \bm v +\nabla p
\nonumber \\
= -\alpha \bm v +\nu \nabla^2\bm v +\bm f,
\label{NavierStokes}
\end{eqnarray}
where $p$ is pressure, $\alpha$ is the bottom friction coefficient, $\nu$ is the kinematic viscosity coefficient, and $\bm n$ is the unit vector along the first axis. Note that the external force supporting the shear flow is equal to $\alpha \varSigma x_2 \bm n$.

In two dimensions, it is convenient to describe the flow in terms of vorticity. We introduce the vorticity $\varpi$ of the fluctuating flow:
\begin{equation}
\varpi= \mathrm{curl}\, \bm v
\equiv \partial_1 v_2-\partial_2 v_1.
\label{varpi}
\end{equation}
Obviously, $\varpi$ is a scalar (or, more precisely, pseudoscalar) field. The equation controlling evolution of vorticity is derived from Eq. (\ref{NavierStokes}):
\begin{equation}
\partial_t\varpi+ \bm v \nabla \varpi
+\varSigma x_2 \frac{\partial \varpi}{\partial x_1}
= -\alpha \varpi +\nu \nabla^2 \varpi +\phi,
\label{basic}
\end{equation}
where $\phi=\mathrm{curl}\, \bm f\equiv \partial_1 f_2-\partial_2 f_1$.

To close the equation (\ref{basic}) one should restore the velocity field $\bm v$ from the vorticity field $\varpi$. Due to the assumed incompressibility condition $\partial_1 v_1+\partial_2 v_2=0$ it is possible to introduce the stream function $\psi$, related to the velocity components and to the vorticity as
\begin{equation}
v_1=\frac{\partial \psi}{\partial x_2}, \quad
v_2=-\frac{\partial \psi}{\partial x_1}, \quad
\varpi=-\nabla^2 \psi.
\label{stre1}
\end{equation}
To find the stream function $\psi$, it is necessary to solve the Laplace equation $\nabla^2 \psi=-\varpi$. Say, one can exploit the integral representation
\begin{equation}
\psi(\bm r)=-\frac{1}{2\pi}
\int d^2 x\, \varpi(\bm x) \ln |\bm r-\bm x|,
\label{stre2}
\end{equation}
implying space homogeneity and isotropy of the system. After finding $\psi$ one calculates the velocity components in accordance with Eq. (\ref{stre1}).

The external random force $\bm f$ pumps energy to the flow. The average rate of the energy production per unit mass is written as $\langle \bm f \bm v \rangle$. The angular brackets here and below designate an averaged value. In laboratory and numerical experiments, the averaging is performed over time. In our theoretical setup, the averaging is performed over the statistics of the pumping force $\bm f$. Besides the energy production one introduces the enstrophy production. The enstrophy production rate per unit mass is written as the average $\langle \phi \varpi \rangle$, analogously to the energy production rate. Both quantities, $\langle \bm f \bm v \rangle$ and $\langle \phi \varpi \rangle$, are assumed to be homogeneous in space and time in our model.

We assume that the pumping force $\bm f$ is short correlated in time and is zero in average, $\langle \bm f \rangle=0$. Then statistical properties of the external pumping are fully determined by its pair correlation function
\begin{equation}
\langle f_\alpha(t,\bm x) f_\beta(0,\bm y)\rangle
=2\epsilon \delta(t)\delta_{\alpha\beta} \varXi(\bm x -\bm y),
\label{forcecorrel}
\end{equation}
reflecting the assumed homogeneity in space and time. We fix $\varXi(\bm 0)=1$, then the energy production rate is $\langle \bm f \bm v \rangle=\epsilon$. Passing to $\phi=\mathrm{curl}\, \bm f$, one finds from Eq. (\ref{forcecorrel})
\begin{equation}
\langle\phi(t,\bm x)\phi(0,\bm y)\rangle
=-2\epsilon\delta(t) \nabla^2 \varXi(\bm x-\bm y),
\label{basic2}
\end{equation}
where $\nabla$ in Laplacian $\nabla^2$ can be either the derivative over $\bm x$ or over $\bm y$.

Generally, the correlation functions of the fluctuating vorticity are not homogeneous in space due to the presence of the mean shear flow, breaking homogeneity along the second coordinate axis. However,  a shift along the second coordinate axis can be compensated by the corresponding Galilean transformation, as it follows from Eq. (\ref{basic}). The relation (\ref{basic2}) is invariant under a Galilean transformation. We conclude that in our model simultaneous correlation functions of vorticity are homogeneous in space. The same is true for the velocity correlation functions. Particularly, all the single-point moments are independent of the coordinates.

Multiplying the equation (\ref{NavierStokes}) by the fluctuating velocity $\bm v$ and averaging the result, one finds the energy balance
\begin{eqnarray}
\epsilon=\varSigma \langle v_1 v_2 \rangle + \alpha \langle v^2 \rangle
+ \nu \langle (\partial_\alpha \bm v)^2 \rangle.
\label{energybal}
\end{eqnarray}
Multiplying the equation (\ref{basic}) by $\varpi$ and averaging, one finds the enstrophy balance
\begin{equation}
\langle \phi \varpi \rangle
= \alpha \langle \varpi^2 \rangle +\nu \langle (\nabla \varpi)^2 \rangle.
\label{enstrophybal}
\end{equation}
At deriving the relations (\ref{energybal},\ref{enstrophybal}) we omitted all complete derivatives over time and coordinates. They are zero due to stationarity and the independence of the coordinates of all single-point moments.

The random force $\bm f$ is assumed to be characterized by a correlation length $k_f^{-1}$, where $k_f$ is the characteristic wave vector of the force. We assume that the following inequality
\begin{equation}
\varSigma \gg \nu k_f^2,
\label{criterion5}
\end{equation}
is satisfied. The inequality (\ref{criterion5}) means that the shear flow is strong enough to influence essentially the flow fluctuations at the pumping scale. Further we consider the case where the condition
\begin{equation}
\alpha \lesssim \nu k_f^2,
\label{criterion2}
\end{equation}
is satisfied. As it was proposed in Ref. \cite{KL20}, the condition (\ref{criterion2}) has to be satisfied for appearing coherent vortices in two-dimensional turbulence. The opposite case $\alpha \gg \nu k_f^2$ needs a special analysis. Combining the inequality (\ref{criterion5}) with Eq. (\ref{criterion2}), we conclude that the inequality $\varSigma \gg \alpha$ is satisfied in our setup as well.

Our calculations, presented below, are conducted at the assumption $\varSigma>0$. However, all the results derived can be easily extended to negative $\varSigma$. Say, the simple symmetry reasoning show that the third moment $\langle \varpi^3 \rangle$ changes its sign at $\varSigma \to-\varSigma$. As to the pair correlation function $\langle \varpi(t,\bm x) \varpi (0,\bm y)\rangle$, the substitution $\varSigma\to - \varSigma$ means that one should change the sign of the second component $x_2\to-x_2$, $y_2\to -y_2$ without touching $x_1,y_1$. Similar reasoning enables one to establish the transformation laws of any correlation function at the substitution $\varSigma\to - \varSigma$.

\section{Perturbation theory}
\label{sec:perturb}

In the linear approximation the equation (\ref{basic}) is reduced to the equation
\begin{equation}
(\partial_t+\hat{\mathcal L})\varpi=\phi, \quad
\hat{\mathcal L}=
\varSigma x_2 \frac{\partial}{\partial x_1}
+\alpha -\nu \nabla^2,
\label{basic1}
\end{equation}
Solutions of the equation (\ref{basic1}) without pumping $\phi$ were analyzed in Ref. \cite{souzy}. To find the correlation functions of $\varpi$ in the linear approximation, one should solve the equation (\ref{basic1}) for arbitrary $\phi$ and then average the corresponding product over the statistics of $\phi$, determined by Eq. (\ref{basic2}).

To find corrections to the expressions, obtained in the linear approximation, one has to take into account the nonlinear term $\bm v \nabla \varpi$ in the equation (\ref{basic}). Solving the equation recursively, one obtains vorticity $\varpi$ as a series over the pumping force $\bm f$. Averaging the products of vorticity over the statistics of the pumping force in accordance with Eq. (\ref{forcecorrel}) one arrive at the perturbation series in terms of powers of the parameter $\epsilon$. Such series produced by the nonlinear hydrodynamic interaction was first considered in Ref. \cite{Wyld61}. The terms of the series can be represented by Feynman diagrams, therefore the technique is called Wyld diagrammatic technique. Wyld diagrammatic technique can be consistently derived from the representation of correlation functions as functional integrals over the observed variables and auxiliary fields \cite{MSR73}. The integration is performed like in the quantum field theory with the weight $\exp(-{\mathcal I})$, where ${\mathcal I}$ is the effective action. A detailed description of the technique can be found in the review \cite{HRS}.

In our case, the functional integral is written in terms of the vorticity $\varpi$ and the corresponding auxiliary field $\mu$. The effective action ${\mathcal I}$ can be derived from the equation (\ref{basic}) and the relation (\ref{basic2}). It is written as
\begin{eqnarray}
{\mathcal I}={\mathcal I}_2+{\mathcal I}_{int},
\label{gener1}
\end{eqnarray}
where
\begin{eqnarray}
{\mathcal I}_2= \int dt\, d^2x\, \mu (\partial_t+\hat{\mathcal L})\varpi
\nonumber \\
+\epsilon \int dt\, d^2x\, d^2 r\, \nabla^2\varXi(\bm x-\bm r) \mu(t,\bm x) \mu(t,\bm r),
\label{gener2} \\
{\mathcal I}_{int}=
\int dt\, d^2x\, \mu \bm v \nabla \varpi .
\label{generint}
\end{eqnarray}
The velocity $\bm v$ in Eq. (\ref{generint}) is implied to be expressed via the vorticity $\varpi$, see Eqs. (\ref{stre1},\ref{stre2}).

Let us introduce the pair correlation functions. The pair correlation function of the vorticity is written as the following functional integral
\begin{equation}
\langle \varpi(t, \bm x) \varpi (s, \bm y)\rangle
= \int D\varpi\, D\mu\, e^{-{\mathcal I}}
 \varpi(t, \bm x) \varpi (s, \bm y),
 \label{paircofu}
\end{equation}
We introduce also the following pair average
\begin{equation}
\langle \varpi(t, \bm x) \mu (s, \bm y)\rangle
= \int D\varpi\, D\mu\, e^{-{\mathcal I}}
 \varpi(t, \bm x) \mu (s, \bm y),
 \label{greenofu}
\end{equation}
that we call the Green function. The reason is that the correlation function (\ref{greenofu}) determines a response of the system to an additional external force. Note that the pair average $\langle \mu \mu \rangle$ is zero.

Because of the time homogeneity the pair correlation function of vorticity (\ref{paircofu}) and the Green function {\ref{greenofu}) depend solely on the time difference $t-s$. Due to the presence of the shear flow the space homogeneity is broken in our model. Therefore the pair correlation function (\ref{paircofu}) and the Green function {\ref{greenofu}) depend on both space coordinates. More precisely, the space homogeneity is broken in the direction of the second axis. Therefore the pair correlation function (\ref{paircofu}) and the Green function {\ref{greenofu}) depend on both second coordinates, $x_2$ and $y_2$, and on the difference $x_1-y_1$.

Any correlation function of $\varpi,\mu$ can be written as the functional integral analogously to the pair correlation function (\ref{paircofu}) and the Green function {\ref{greenofu}). One can evolve the perturbation theory for the correlation function expanding the weight $\exp(-{\mathcal I})$ in the functional integral over the third order term ${\mathcal I}_{int}$ (\ref{generint}) and calculating the resulting Gaussian functional integrals. The integrals are expressed in terms of the ``bare'' correlation functions determined by the quadratic term (\ref{gener2}) in the effective action:
\begin{eqnarray}
\langle \varpi(t,\bm x) \mu(0,\bm y)\rangle_0
=\int {\mathcal D}\varpi {\mathcal D}\mu\, e^{-{\mathcal I}_2}
\varpi(t,\bm x) \mu(0,\bm y),
\label{gener3} \\
\langle \varpi(t,\bm x) \varpi(0,\bm y)\rangle_0
=\int {\mathcal D}\varpi {\mathcal D}\mu\, e^{-{\mathcal I}_2}
\varpi(t,\bm x) \varpi(0,\bm y).
\label{genar3}
\end{eqnarray}
The average of the type $\langle \varpi \dots \mu \dots\rangle_0$ is determined by Wick theorem \cite{Wick50} and it is equal to the sum of products of the pair averages (\ref{gener3},\ref{genar3}) organized by all possible pairings.

The procedure enables one to construct the perturbation series in the spirit of quantum field theory, see, e.g., \cite{qufield}. To use the perturbation series one should know the ``bare'' correlation functions (\ref{gener3},\ref{genar3}). We proceed to calculating the correlation functions.

\subsection{``Bare'' correlation functions}

The ``bare'' correlation functions (\ref{gener3},\ref{genar3}) are expressed in terms of the Gaussian integrals and can be easily found explicitly. The expression (\ref{gener2}) lead to the following equation for the ``bare'' Green function
\begin{equation}
(\partial_t+\hat{\mathcal L})
\langle \varpi(t,\bm x) \mu(0,\bm y)\rangle_0
=\delta(t)\delta(\bm x-\bm y).
\label{greenf}
\end{equation}
Remind that any Green function is zero at negative times due to causality. Therefore $\langle \varpi(t,\bm x) \mu(0,\bm y)\rangle$ is zero at $t<0$. The expression for the ``bare'' pair correlation function can be derived using Eq. (\ref{genar3}):
\begin{eqnarray}
\langle \varpi(t,\bm x) \varpi(0,\bm y)\rangle_0
=-2\epsilon\int d\tau \int d^2r d^2 z
\nabla^2\varXi(\bm r-\bm z)
\nonumber \\
\langle \varpi(t,\bm x) \mu(\tau,\bm r)\rangle_0
\langle \varpi(0,\bm y) \mu(\tau,\bm z)\rangle_0 . \quad
\label{genar4}
\end{eqnarray}
Note that $\tau<0$ and $\tau<t$ in the integral (\ref{genar4}).

Let us pass to Fourier transforms of the pair correlation function (\ref{genar4}) and the Green function {\ref{greenf}). Since they depend on both coordinates, the transform includes two integrals. We use the following definitions:
\begin{eqnarray}
\langle \varpi(t,\bm x) \mu(0,\bm y)\rangle_0
=\int \frac{d^2k\, d^2q}{(2\pi)^4}
e^{i\bm k \bm x-i\bm q \bm y}
{\mathcal G}(t,\bm k,\bm q),
\label{fouri1} \\
\langle \varpi(t,\bm x) \varpi(0,\bm y)\rangle_0
=\int \frac{d^2k\, d^2q}{(2\pi)^4}
e^{i\bm k \bm x+i\bm q \bm y}
{\mathcal F}(t,\bm k,\bm q).
\label{fouri2}
\end{eqnarray}
The pair correlation function (\ref{fouri2}) is invariant under the substitution $t\to -t$, $\bm x \leftrightarrow \bm y$ or $\bm k \leftrightarrow \bm q$.

For the Fourier transform ${\mathcal G}(t,\bm k,\bm q)$ (\ref{fouri1}) we derive from Eq. (\ref{greenf}) the following differential equation
\begin{eqnarray}
\left(\frac{\partial}{\partial t} - \varSigma k_1\frac{\partial}{\partial k_2}
+\alpha+\nu k_1^2+\nu k_2^2\right){\mathcal G}(t,\bm k,\bm q)
\nonumber \\
=(2\pi)^2\delta(t)\delta(\bm k-\bm q).
\label{gref1}
\end{eqnarray}
Since the equation (\ref{gref1}) is of the first order, it can be easily solved by the method of characteristics to obtain
\begin{eqnarray}
{\mathcal G}(t,\bm k,\bm q)
=(2\pi)^2 \theta(t) \delta(k_1-q_1)
\nonumber \\
\delta\left(k_2-q_2+\varSigma k_1t\right)
G(t,\bm q),
\label{gref2} \\
G(t,\bm q)=
\exp\left(-\alpha t-\nu \bm q^2 t
\vphantom{\frac{1}{3}}\right.
\nonumber \\ \left.
+\nu\varSigma q_2 q_1 t^2-\frac{1}{3}\nu \varSigma^2 q_1^2 t^3
\right),
\label{zerot1}
\end{eqnarray}
where $\theta(t)$ is Heaviside step function, $\theta(t)=1$ if $t>0$ and $\theta(t)=0$ if $t<0$. The function $\theta(t)$ in Eq. (\ref{zerot1}) reflects causality: ${\mathcal G}=0$ if $t<0$.

There is the general property of Green functions
\begin{equation}
{\mathcal G}(t+\tau,\bm k, \bm q)=
\int \frac{d^2p}{(2\pi)^2}
{\mathcal G}(t,\bm k, \bm p)
{\mathcal G}(\tau,\bm p, \bm q),
\label{genera1}
\end{equation}
where $t>0,\tau>0$. Using Eq. (\ref{genera1}), one finds
\begin{equation}
G(t+\tau,\bm q)
=G(t,q_1,q_2-\varSigma q_1 \tau) G(\tau,\bm q).
\label{convolution}
\end{equation}
Of course, one can directly check the relation (\ref{convolution}), using the expression (\ref{zerot1}). Note also the expression
\begin{eqnarray}
\int \frac{d^2 k}{(2\pi)^2} \exp(i\bm k \bm x)
{\mathcal G}(t,\bm k,\bm q)
\nonumber \\
=\theta(t) \exp\left[iq_1 x_1 +i(q_2-\varSigma q_1 t)x_2\right]
G(t,\bm q),
\label{mixst}
\end{eqnarray}
for the partial Fourier transform of Eq. (\ref{gref2}).

We obtain from Eqs. (\ref{genar4},\ref{fouri1},\ref{fouri2})
\begin{eqnarray}
{\mathcal F}(t, \bm k, \bm q)=
2\epsilon\int d\tau \int \frac{d^2 p}{(2\pi)^2}
p^2 \tilde\varXi(\bm p)
\nonumber \\
{\mathcal G}(t+\tau,\bm k,\bm p)
{\mathcal G}(\tau,\bm q,-\bm p).
\label{fouri4}
\end{eqnarray}
Here
\begin{equation}
\tilde\varXi(\bm k)=\int d^2 x\, \exp(-i\bm k \bm x) \varXi(\bm x),
\label{tildexi}
\end{equation}
is spacial Fourier transform of $\varXi(\bm r)$.

Substituting the expression (\ref{gref2}) into Eq. (\ref{fouri4}), one finds
\begin{eqnarray}
{\mathcal F}(t, \bm k, \bm q)=
2\epsilon (2\pi)^2 \delta(k_1+q_1)\delta(k_2+q_2+\varSigma k_1 t)
\nonumber \\
\int d\tau\, \theta(\tau) \theta(t+\tau)
\bm p^2 \tilde \varXi(\bm p) G(t+\tau, \bm p) G(\tau,\bm p)
\label{reduc1}
\end{eqnarray}
where
\begin{equation}
p_1=k_1=-q_1, \quad
p_2=k_2+\varSigma (t+\tau) k_1
=-q_2- \varSigma \tau q_1.
\nonumber
\end{equation}
For the simultaneous correlation function one obtains
\begin{eqnarray}
{\mathcal F}(0,\bm k, \bm q)=
(2\pi)^2 \delta(\bm k+\bm q) F(\bm q),
\label{simul} \\
F(\bm k)=2\epsilon \int_0^\infty d\tau\, \bm p^2 \tilde\varXi(\bm p)
G^2(\tau,\bm p),
\label{simul2}
\end{eqnarray}
where $p_1=k_1$, $p_2=k_2+\varSigma \tau k_1$. Using Eq. (\ref{convolution}) one finds from Eq. (\ref{reduc1})
\begin{eqnarray}
{\mathcal F}(t,\bm k, \bm q)=
(2\pi)^2 \delta(k_1+q_1)
\delta(k_2+q_2 +\varSigma t k_1)
\nonumber \\
G(|t|,k_1,k_2+\varSigma t k_1) F(\bm q).
\label{simul4}
\end{eqnarray}
Of course, for $t\to 0$ we return to Eq. (\ref{simul}).

\subsection{Some ``bare'' quantities}

Let us examine the bare contribution to the non-diagonal component of Reynolds stress
\begin{eqnarray}
\langle v_1(0,\bm 0) v_2(0,\bm 0) \rangle_0
=\int \frac{d^2 k\, d^2 q}{(2\pi)^4}
\frac{k_1q_2}{k^2 q^2}
{\mathcal F}(0, \bm k, \bm q),
\nonumber
\end{eqnarray}
where the factor at $\mathcal F$ in the integrand is obtained from Eq. (\ref{stre1}). Substituting here the expressions (\ref{simul},\ref{simul2}) one obtains
\begin{eqnarray}
\langle v_1(0,\bm 0) v_2(0,\bm 0) \rangle_0
\nonumber \\
=2\epsilon \int_0^\infty d\tau \int \frac{d^2p}{(2\pi)^2} \bm p^2 \tilde \varXi(\bm p)
[{ G}(\tau,\bm p)]^2
\nonumber \\
\frac{p_1 (-p_2+\varSigma p_1 \tau)}{[p_1^2+ (-p_2+\varSigma p_1 \tau)^2]^2},
\label{reyn1}
\end{eqnarray}
where ${ G}(\tau,\bm p)$ is determined by Eq. (\ref{zerot1}). Here $p_1\sim p_2\sim k_f$ and the characteristic $\tau$ is determined by the denominator, that is $\tau\sim \varSigma^{-1}$, and ${ G}(\tau,\bm p)$ can be substituted by unity. Taking then the integral over $\tau$,
\begin{eqnarray}
\int_0^\infty d\tau \frac{q (-p+ q \varSigma\tau)}{[q^2+ (p- q \varSigma\tau)^2]^2}
=\frac{1}{2\varSigma(q^2+p^2)},
\nonumber
\end{eqnarray}
one finds the result \cite{KL15,kolokolov2016structure}
\begin{eqnarray}
\langle v_1(0,\bm 0) v_2(0,\bm 0) \rangle_0
=\frac{\epsilon}{\varSigma} \int \frac{d^2p}{(2\pi)^2} \tilde \varXi(\bm p)
=\frac{\epsilon}{\varSigma},
\label{reyn2}
\end{eqnarray}
since $\varXi(0)=1$ by definition.

The expressions (\ref{simul},\ref{simul2}) enable one to evaluate the bare value of the second moment of the vorticity:
\begin{eqnarray}
\langle \varpi^2 \rangle_0
=2\epsilon \int_0^\infty d\tau\,
\int \frac{d^2 p}{(2\pi)^2}
p^2 \tilde\varXi(\bm p) G^2(\tau,\bm p).
\nonumber
\end{eqnarray}
The values of the components of the wave vector $\bm p$ can be estimated as $k_f$. Therefore in the main approximation we can keep solely the last term in the exponent in Eq. (\ref{zerot1}):
\begin{equation}
G(\tau,\bm p) \to
\exp\left(-\frac{1}{3}\nu \varSigma^2 p_1^2 \tau^3 \right).
\label{substi}
\end{equation}
Substituting then $p_1\sim p_2 \sim k_f$, we end up with the estimate
\begin{equation}
\langle \varpi^2 \rangle_0
\sim \epsilon k_f^2 \tau_\star , \quad
\tau_\star =\left(\varSigma^2 \nu k_f^2\right)^{-1/3}.
\label{secondm}
\end{equation}
Note that $\varSigma \gg \tau_\star^{-1}\gg \nu k_f^2$. The inequalities are explained by the condition (\ref{criterion5}). Note also that $\alpha \tau_\star \ll 1$. The inequality is explained by the same condition (\ref{criterion5}) and the condition (\ref{criterion2}).

One can find the factor in the law (\ref{secondm}) for a particular function $\tilde \varXi(\bm p)$. Say, one can take
\begin{equation}
\tilde \varXi(\bm p)=\frac{2\pi}{k_f^2}
\exp\left(-\frac{p^2}{2 k_f^2}\right),
\label{gausspu}
\end{equation}
corresponding to $\varXi(\bm x)= \exp(-k_f^2 x^2/2)$. Then one finds \cite{KLT23b}
\begin{equation}
\langle \varpi^2 \rangle_0
=\frac{2^{10/3}}{3^{2/3}\sqrt\pi}
\Gamma(1/3)\Gamma(7/6)
\epsilon k_f^2 \tau_\star.
\nonumber
\end{equation}
The result of the calculation confirms the general estimate (\ref{secondm}).

Analogously, the bare pair correlation function can be evaluated \cite{KLT23b}. The simultaneous correlation function (\ref{simul}) is determined by $F(\bm k)$ that is given by the integral (\ref{simul2}). If $k_f\gtrsim k_1\gg \nu k_f^3/\varSigma$, $k_f\gtrsim k_2$ then the characteristic time in the integral (\ref{simul2}) is determined by $k_1$, see Eq. (\ref{substi}), and we obtain
\begin{equation}
F\sim \epsilon \tau_\star k^2 k_f^{-4/3} k_1^{-2/3}.
\label{occup}
\end{equation}
Thus for $k_1,k_2\sim k_f$ we have $F\sim \epsilon \tau_\star$ and for $k_1\sim \nu k_f^3/\varSigma, k_2\sim k_f$ we have $F\sim \epsilon/(\nu k_f^2)$. The expression (\ref{occup}) gives the universal scaling behavior between the limit cases.

One can obtain a convenient expression for $F(\bm k)$ using the particular form (\ref{gausspu}) of the pumping correlations. If $k_f\gtrsim k_1\gg \nu k_f^3/\varSigma$ then
\begin{equation}
F(\bm k)=
\left(\frac{3}{2}\right)^{1/3}\Gamma\left(\frac{4}{3}\right)
\frac{4\pi\epsilon \tau_\star k^2}{k_f^{4/3} k_1^{2/3}}
\exp\left(-\frac{k^2}{2 k_f^2}\right),
\label{gausspc}
\end{equation}
in accordance with the law (\ref{occup}). Fourier transform of the expression (\ref{gausspc}) enables one to restore the properties of the simultaneous pair correlation function in real space examined in Ref. \cite{KLT23b}.

\subsection{Interaction corrections}

As we explained, interaction corrections to the bare values of the correlation functions has to be calculated in the framework of the perturbation series. It is constructed by the expansion of the factor $\exp({\mathcal I}_2+{\mathcal I}_{int})$ in the series over ${\mathcal I}_{int}$ (\ref{generint}) in the corresponding functional integral. Each term of the expansion can be found analytically, using Wick theorem \cite{HRS}. It is a multiple integral over times and wave vectors of some expression determined by the ``bare'' pair correlation functions, Green functions and factors corresponding to converting $\varpi \to \bm v$. Note that to increase the order of the perturbation series by unity, one has to take into account two additional terms of the expansion over ${\mathcal I}_{int}$.

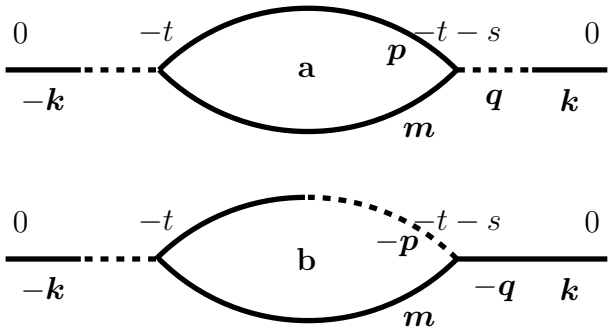
\begin{figure}
\begin{tikzpicture} <br>
\draw[line width=2.0pt,dashed] (7,7.5) -- (8,7.5);
\draw[line width=2.0pt](8,7.5) -- (9,7.5);
\draw[line width=2.0pt,dashed] (3,7.5) -- (2,7.5);
\draw[line width=2.0pt](2,7.5) -- (1,7.5);
\node at (5,7.5) {\large \bf a};
\node at (1.2,8) {\large $0$};
\node at (8.8,8) {\large $0$};
\node at (3,8) {\large $-t$};
\node at (7,8) {\large $-t-s$};
\node at (1.5,7.1) {\large $-\bm k$};
\node at (8.5,7.1) {\large $\bm k$};
\node at (7.5,7.1) {\large $\bm q$};
\node at (6.2,7.7) {\large $\bm p$};
\node at (6.5,6.7) {\large $\bm m$};
\node at (6.5,6) {\phantom .};
\draw[line width=2.0pt] (7,7.5) arc (-45:-135:2.8);
\draw[line width=2.0pt] (7,7.5) arc (45:135:2.8);
\end{tikzpicture}
\begin{tikzpicture} <br>
\draw[line width=2.0pt](7,7.5) -- (9,7.5);
\draw[line width=2.0pt,dashed] (3,7.5) -- (2,7.5);
\draw[line width=2.0pt](2,7.5) -- (1,7.5);
\node at (5,7.5) {\large \bf b};
\node at (1.2,8) {\large $0$};
\node at (8.8,8) {\large $0$};
\node at (3,8) {\large $-t$};
\node at (7,8) {\large $-t-s$};
\node at (1.5,7.1) {\large $-\bm k$};
\node at (8.5,7.1) {\large $\bm k$};
\node at (7.5,7.1) {\large $-\bm q$};
\node at (6.2,7.7) {\large $-\bm p$};
\node at (6.5,6.7) {\large $\bm m$};
\draw[line width=2.0pt] (7,7.5) arc (-45:-135:2.8);
\draw[line width=2.0pt,dashed] (7,7.5) arc (45:90:2.8);
\draw[line width=2.0pt] (3,7.5) arc (135:90:2.8);
\end{tikzpicture}
\caption{Feynman diagrams representing first corrections to the simultaneous pair correlation function.}
\label{fig:paircorr}
\end{figure}

The terms of the perturbation series can be represented by Feynman diagrams. The diagrams determined the first order correction to the pair correlation function of vorticity are depicted in Fig. \ref{fig:paircorr}. Higher order corrections correspond to more complicated diagrams. An example of more complicated diagram determining the second order correction to the pair correlation function of vorticity is depicted in Fig. \ref{fig:paircorr2}. The order of the perturbation series corresponds to the number of loops of the diagrams. Say, the first order correction to the pair correlation function of vorticity is determined by the one-loop diagrams, see Fig. \ref{fig:paircorr}, and the second order correction to the pair correlation function of vorticity is determined by the two-loop diagrams, see Fig. \ref{fig:paircorr2}.

All lines on the diagrams are thought to be consistent of two segments, solid segments correspond to the field $\varpi$ whereas dashed segments correspond to the field $\mu$. Thus, a combined solid-dashed line designates the Green function ${\mathcal G}(t,\bm k,\bm q)$ and a solid line designate the pair correlation function ${\mathcal F}(t,\bm k,\bm q)$. Two solid segments and one dashed segment are attached to each vertex in accordance with the structure of ${\mathcal I}_{int}$ (\ref{generint}). The factor corresponding to each vertex is
\begin{equation}
\frac{1}{2}\left(\frac{1}{q^2}-\frac{1}{k^2}\right)
(q_2 k_1 -q_1k_2),
\label{turb1}
\end{equation}
where $\bm k, \bm q$ are wave vectors of the solid segments attached to the vertex.

To construct the analytical expression, corresponding to a given diagram, one should fix the corresponding combinatorial factor, take the product of the Green functions, of the pair correlation functions and of the vertex factor and integrate the result over ``internal'' wave vectors and times. The integration should be performed at the condition of the wave vector conservation at each vertex: the sum of the wave vectors of three segments attached to the vertex has to be zero. It is interesting that the corrections to $\mathcal G$, $\mathcal F$ reproduce the structure of the bare correlation functions (\ref{gref2},\ref{simul4}) with the same $\delta$-functions. Thus, one can extract from the diagrams corrections to the functions $G(t,\bm k)$, $F(\bm k)$. The property is a consequence of assumed short correlation in time of pumping. That is why we have chosen the model, since the property simplifies essentially concrete calculations.

The perturbation series is presented by a set of diagrams with different number of loops. Enhancement by unity of the number of the loops means adding one $F$-line, one $G$-line and two vertices to the diagram. Note that $F$ contains $\epsilon$ as a factor, see Eq. (\ref{simul2}), whereas $G$ and the factor (\ref{turb1}), corresponding to a vertex, do not. Therefore one can say, that the perturbation series is a series over $\epsilon$. The dimensionless parameter controlling the perturbation series is
\begin{equation}
\beta= \frac{\epsilon}{\varSigma^2 \nu}.
\label{turb4}
\end{equation}
For different objects, the parameter (\ref{turb4}) can be corrected by factors depending on the dimensionless quantities $\nu k_f^2/\varSigma$, $k_1/k_f$, $k_2/k_f$. Any case, the parameter (\ref{turb4}) should be small for validity of the perturbation series.

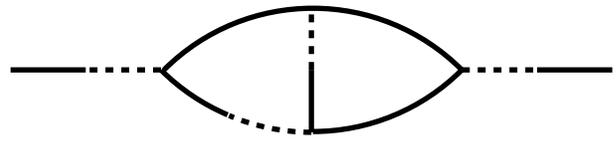
\begin{figure}
\begin{tikzpicture} <br>
\draw[line width=2.0pt,dashed] (7,7.5) -- (8,7.5);
\draw[line width=2.0pt](8,7.5) -- (9,7.5);
\draw[line width=2.0pt,dashed] (3,7.5) -- (2,7.5);
\draw[line width=2.0pt](2,7.5) -- (1,7.5);
\draw[line width=2.0pt](5,6.67) -- (5,7.5);
\draw[line width=2.0pt,dashed] (5,7.5) -- (5,8.33);
\draw[line width=2.0pt] (7,7.5) arc (-45:-90:2.8);
\draw[line width=2.0pt] (3,7.5) arc (-135:-113:2.8);
\draw[line width=2.0pt,dashed] (5,6.67) arc (-90:-113:2.8);
\draw[line width=2.0pt] (7,7.5) arc (45:135:2.8);
\end{tikzpicture}
\caption{Two-loop diagram representing a second-order correction to the pair correlation function}
\label{fig:paircorr2}
\end{figure}

Note that zero contribution to the third moment of vorticity $\langle \varpi^3 \rangle$ is zero. The first non-vanishing contribution to $\langle \varpi^3 \rangle$ appears in the first order in the perturbation theory. Let us stress that it is determined by a ``tree'' diagram, where loops are absent. Thus, the contribution needs a special analysis made in Ref.  \cite{KLT23b}. The result is
\begin{equation}
\langle \varpi^3 \rangle \sim \frac{\epsilon^2 k_f^2}{\varSigma^2 \nu},
\label{thirdm}
\end{equation}
upto a logarithmic factor. The quantity (\ref{thirdm}) satisfies the relation
\begin{equation}
\langle \varpi^3 \rangle^2\sim \beta
(\langle \varpi^2 \rangle_0)^{3},
\label{thirdm2}
\end{equation}
see Eq. (\ref{secondm}). The factor $\beta$ in Eq. (\ref{thirdm2}) is a manifestation of the fact that the main contribution to the third moment of vorticity appears in the first order of the perturbation series.

\section{Corrections to the pair correlation function of vorticity}
\label{sec:corrections}

We proceed to calculating corrections to the pair correlation function. The calculation reveals the peculiarities of the perturbation series and demonstrates some universal features, that can be used for evaluating more complicated objects. We concentrate on analyzing the simultaneous correlation function, depending solely on the coordinate difference. The correction is written as
\begin{equation}
\delta\langle \varpi(\bm x) \varpi(\bm 0) \rangle
= \int\frac{d^2 k}{(2\pi)^2} \delta F(\bm k),
\label{corrfouri}
\end{equation}
where $\delta F(\bm k)$ is the correction to the function $F(\bm k)$ introduced by Eq. (\ref{simul4}).

In the first order of the perturbation series the correction $\delta F(\bm k)$ is determined by the diagrams depicted in Fig. \ref{fig:paircorr}. Let us stress that we deal with nonsimultaneous correlation functions inside the diagrams. Therefore one should use the general functions ${\mathcal G}(t,\bm k,\bm q)$, ${\mathcal F}(t,\bm k,\bm q)$, see Eqs. (\ref{gref2},\ref{zerot1},\ref{simul4}). The analytical expressions for the contributions to $\delta F(\bm k)$, corresponding to the diagrams {\bf a} and {\bf b} in Fig. \ref{fig:paircorr}, are
\begin{eqnarray}
F_{\mathrm a}(\bm k)=\int^{\infty}_0 dt \int_{0}^\infty ds \int \frac{d^2 m}{(2\pi)^2}
\nonumber \\
G(t,-k_1,-k_2-\varSigma t k_1) G(s+t, \bm q)
\nonumber \\
V(\bm m,\bm p) G(s,\bm p)
G(s,\bm m) F(\bm m) F(\bm p)
\nonumber \\
\left(\frac{1}{\bm m^2}-\frac{1}{\bm p^2}\right)
(m_2p_1-m_1 p_2),
\label{corrpaia}
\end{eqnarray}
and
\begin{eqnarray}
F_{\mathrm b}(\bm k)=-2\int^{\infty}_0 dt \int_{0}^\infty ds \int \frac{d^2 m}{(2\pi)^2}
\nonumber  \\
G(t,-k_1,-k_2-\varSigma t k_1) G(s+t, \bm q)
\nonumber \\
V(\bm m,\bm p) G(s,\bm p)
G(s,\bm m) F(\bm m)  F(\bm q)
\nonumber \\
\left(\frac{1}{\bm m^2}-\frac{1}{{\bm q}^2}\right)
(m_2q_1-m_1 q_2),
\label{corrpaib}
\end{eqnarray}
where $q_1=k_1$, $q_2=k_2+\varSigma (t+s) k_1$, $p_1=k_1-m_1$, $p_2=k_2-m_2 +\varSigma (t+s) k_1$, and $V$ is the factor (\ref{turb1}) corresponding to the left vertices:
\begin{eqnarray}
V(\bm m,\bm p)=\frac{1}{2}\left\{\frac{1}{m_1^2+(m_2-\varSigma s m_1)^2}-
\right. \nonumber \\ \left.
\frac{1}{p_1^2+(p_2-\varSigma s p_1)^2}\right\}
(m_2p_1-m_1 p_2).
\label{turb11}
\end{eqnarray}
The expressions are written in accordance with the general rules of reading diagrams.

The integrand in Eq. (\ref{corrpaia}) is invariant under the permutation $\bm m \leftrightarrow \bm p$. Therefore one can substitute
\begin{equation}
\frac{1}{m^2}-\frac{1}{p^2}\to \frac{2}{m^2}.
\nonumber
\end{equation}
After the substitution we find
\begin{eqnarray}
\delta F(\bm k) =F_{\mathrm a}+F_{\mathrm b}
=2\int^{\infty}_0 dt \int_{0}^\infty ds \int \frac{d^2 m}{(2\pi)^2}
\nonumber  \\
G(t,-k_1,-k_2-\varSigma t k_1) G(s+t, \bm q)
\nonumber \\
V(\bm m,\bm p) G(s,\bm p)
G(s,\bm m) F(\bm m)
\nonumber \\
\left[\frac{F(\bm q)}{\bm q^2}+\frac{F(\bm p)-F(\bm q)}{{\bm m}^2}\right]
(m_2q_1-m_1 q_2),
\label{corrpai1}
\end{eqnarray}
where $\bm p=\bm q-\bm m$.

\subsection{Intermediate region of wave vectors}

Let us consider the case
\begin{equation}
\frac{\nu k_f^2}{\varSigma}
\ll k_1 \ll k_f, \quad k_2\sim k_f,
\label{corrpai2}
\end{equation}
corresponding to the universal bare behavior (\ref{occup}). There are some regions of the integration over $m_1$ in the integral (\ref{corrpai1}) that have to be examined to extract the main contribution to the correction $\delta F$.

Let us examine the region of the integration $m_1\sim m_2\sim k_f$. Then $m_2q_1-m_1 q_2\approx m_1 q_2$, $p_1\approx -m_1$, and
\begin{eqnarray}
\delta F(\bm k)
=\int^{\infty}_0 dt \int_{0}^\infty ds \int \frac{d^2 m}{(2\pi)^2}
\nonumber  \\
G(t,-k_1,-k_2-\varSigma t k_1) G(s+t, \bm q)
\nonumber \\
m_1^2 q_2^2\left\{\frac{1}{m_1^2+(m_2-\varSigma s m_1)^2}
\right. \nonumber \\ \left.
-\frac{1}{p_1^2+(p_2-\varSigma s p_1)^2}\right\}
G(s,\bm p) G(s,\bm m)
\nonumber \\
F(\bm m)
\left[\frac{F(\bm q)}{\bm q^2}+\frac{F(\bm p)-F(\bm q)}{{\bm m}^2}\right].
\nonumber
\end{eqnarray}
Here the integration over $s$ is determined by the denominators and $s\sim \varSigma ^{-1}$. Thus dependencies on $s$ everywhere except the denominators can be neglected: in the Green functions $s\sim \varSigma ^{-1}$ gives small corrections and $s$ enters to $q_2$, $p_2$ with the small factor $k_1$. Since $s$ enters the denominator via the factor $m_1 s$ and all other factors are functions of $m_1^2$, we may integrate over $m_1$ from $0$ to $\infty$ and over $s$ from $-\infty$ to $+\infty$. Next,
\begin{equation}
\int_{-\infty}^{\infty} ds
\frac{1}{m_1^2+(m_2-\varSigma s m_1)^2}
=\frac{\pi}{\varSigma m_1^2}.
\nonumber
\end{equation}
Thus, the integrals over $s$ of the difference of the terms depending on $\bm m$ and $\bm p$ cancel each other. Thus, the main contribution to the correction $\delta F$ from the region of the integration $m_1\sim m_2\sim k_f$ is absent.

Therefore the main contribution to the integral (\ref{corrpai1}) is gained from the region $m_1\sim k_1$ and $m_2\sim k_f$. Then the characteristic times are determined by $F(\bm p)$, $F(\bm q)$, that is
\begin{equation}
s,t \sim \frac{k_f}{\varSigma k_1}, \quad
\varSigma^{-1}\ll s,t \ll (\nu k_f^2)^{-1} .
\label{corrpai3}
\end{equation}
The estimates (\ref{corrpai3}) explain the relation $m_1\sim k_1$ since $m_1$ is determined by the denominators of $V(\bm m, \bm p)$ (\ref{turb11}). The estimates (\ref{corrpai3}) mean that all the Green functions in Eq. (\ref{corrpai1}) can be substituted by unity. Thus we arrive at
\begin{eqnarray}
\delta F(\bm k)
=2\int^{\infty}_0 dt \int_{0}^\infty ds \int \frac{d^2 m}{(2\pi)^2}
V(\bm m,\bm p)
\nonumber \\
F(\bm m)
\left[\frac{F(\bm q)}{\bm q^2}+\frac{F(\bm p)-F(\bm q)}{{\bm m}^2}\right]
(m_2q_1-m_1 q_2),
\label{corrpai4}
\end{eqnarray}
where $\bm p=\bm q-\bm m$.

The inequalities $m_1\ll m_2$ and $p_1 \ll p_2$ enables one to substitute
\begin{eqnarray}
V(\bm m,\bm p)\to
\frac{\pi}{2}(m_2 p_1-m_1 p_2)
\nonumber \\
\left[\frac{1}{|m_1|}\delta(m_2-\varSigma s m_1)
-\frac{1}{|p_1|}\delta(p_2-\varSigma s p_1) \right],
\label{deltaf}
\end{eqnarray}
as it follows from Eq. (\ref{turb11}). Then we find from Eqs. (\ref{turb11},\ref{corrpai4})
\begin{eqnarray}
\delta F(\bm k)
=\int^{\infty}_0 dt \int_{0}^\infty ds \int \frac{d^2 m}{4\pi}
\nonumber \\
\left\{\frac{1}{|m_1|}
\delta\left(\varSigma s m_1-{m_2}\right)-
\frac{1}{|p_1|}
\delta\left(\varSigma s p_1-p_2\right)
\right\}
\nonumber \\
F(\bm m)F(\bm q)\left(\frac{1}{q^2}-\frac{1}{m^2}\right)
(m_2q_1-m_1 q_2)^2.
\label{corrpoi1}
\end{eqnarray}
One can rewrite the expression (\ref{corrpoi1}) as
\begin{eqnarray}
\delta F(\bm k)
=\iint^{\infty}_0 dt\, ds
\int_{-\infty}^{+\infty} \frac{d m_1}{4\pi}
 |m_1|(k_2+\varSigma t k_1)^2
\nonumber \\
F(\bm q) \left[F(\bm m)\left(\frac{1}{q^2}-\frac{1}{m^2}\right)
-F(\bm p)\left(\frac{1}{q^2}-\frac{1}{p^2}\right)\right],
\label{corrpoi2}
\end{eqnarray}
where
\begin{eqnarray}
m_2=\varSigma s m_1, \quad
q_1=k_1, \quad
q_2=k_2+\varSigma (t+s) k_1.
\nonumber \\
p_1=k_1-m_1, \quad
p_2=k_2 -\varSigma s m_1+\varSigma (t+s) k_1,
\nonumber
\end{eqnarray}
The signs of $k_1,k_2$ are arbitrary here.

Let us check that the integral (\ref{corrpoi2}) converges at $m_1\gg k_1$. Then in the main approximation
\begin{equation}
q_2=k_2+\varSigma t k_1, \
p_1=-m_1, \ p_2=q_2 - m_2.
\nonumber
\end{equation}
Passing from the integration over $s$ to the integration over $m_2=\varSigma s m_1$, we find
\begin{eqnarray}
\delta F(\bm k)
=\frac{1}{4\pi}\iint^{\infty}_0 dt\, dm_1
\int_{-\infty}^{+\infty} dm_2\, q_2^2 F(\bm q)
\nonumber \\
\left[F(\bm m)\left(\frac{1}{\bm q^2}-\frac{1}{\bm m^2}\right)
-F(\bm p)\left(\frac{1}{\bm q^2}-\frac{1}{\bm p^2}\right)\right].
\nonumber
\end{eqnarray}
Shifting the integral over $m_2$ in the second term in the square brackets, we arrive at zero main contribution to the integral. Next terms of the expansion of the integrand in Eq. (\ref{corrpoi2}) over the parameter $k_1/m_1$ give converging integrals.

The expression (\ref{corrpoi2}) enables one to evaluate the correction $\delta F$ in the region (\ref{corrpai2}). Combining the estimates $m_1\sim k_1$, (\ref{corrpai3}), and the relation (\ref{occup}), one finds
\begin{equation}
\delta F(\bm k)
\sim \beta \left(\frac{\nu k_f^3}{\varSigma k_1}\right)^{2/3}F(\bm k) ,
\label{estimcorr}
\end{equation}
where $\beta$ is defined by Eq. (\ref{turb4}). The factor $\nu k_f^3/(\varSigma k_1)$ is small for the considered region (\ref{corrpai2}). It becomes of order unity at $k_1\sim \nu k_f^3/\varSigma$, where the relative correction to $F$ (\ref{estimcorr}) is estimated as $\beta$ (\ref{turb4}).

Let us find the correction $\delta F$ using the particular function (\ref{gausspc}). Since $\delta F(\bm k)$ is symmetric under $\bm k\to -\bm k$, without loss of generality, we assume $k_1>0$ whereas $k_2$ is assumed to have an arbitrary sign. We substitute the expression (\ref{gausspc}) into Eq. (\ref{corrpoi2}) and pass to the dimensionless variables $\kappa, \tau. \sigma, \mu$ in accordance with
\begin{eqnarray}
m_1=k_1 \mu, \  t=\frac{k_f}{\varSigma k_1}\tau, \
s=\frac{k_f}{\varSigma k_1}\sigma, \ k_2=k_f \kappa,
\nonumber \\
p_1=k_1(1-\mu), \ q_2=k_f(\kappa+\tau+\sigma),
\nonumber \\
m_2=k_f \sigma \mu, \
 p_2=k_f(\kappa-\sigma \mu+\tau+\sigma).
\end{eqnarray}
Then we arrive at
\begin{equation}
\frac{\delta F(\bm k)}{F(\bm k)}=
\beta \left(\frac{\nu k_f^3}{\varSigma k_1}\right)^{2/3}
A\left(\frac{k_2}{k_f}\right),
\label{corrpoi3}
\end{equation}
where
\begin{eqnarray}
A(\kappa)=
\left(\frac{3}{2}\right)^{1/3}\Gamma\left(\frac{4}{3}\right)
\frac{e^{\kappa^2/2}}{\kappa^2}
\iiint_0^\infty d\tau\, d\sigma\, d\mu
\nonumber \\
\mu (\kappa +\tau)^2
 \exp\left[-\frac{(\kappa+\tau+\sigma)^2}{2}\right]
\nonumber \\
\left\{2\frac{\mu^2\sigma^2-(\kappa+\tau+\sigma)^2}{\mu^{2/3}}
\exp\left[-\frac{\mu^2 \sigma^2}{2}\right]
 \right.
\nonumber \\
-\left[ (\kappa-\sigma \mu+\tau+\sigma)^2-(\kappa+\tau+\sigma)^2\right]
\nonumber \\
\frac{1}{|1-\mu|^{2/3}}
\exp\left[-\frac{(\kappa-\sigma \mu+\tau+\sigma)^2}{2}\right]
\nonumber \\
-\left[ (\kappa+\sigma \mu+\tau+\sigma)^2-(\kappa+\tau+\sigma)^2\right]
\nonumber \\ \left.
\frac{1}{(1+\mu)^{2/3}}
\exp\left[-\frac{(\kappa+\sigma \mu+\tau+\sigma)^2}{2}\right]
\right\}.
\label{intcorrpair}
\end{eqnarray}
The integral (\ref{intcorrpair}) found numerically is plotted in Fig. \ref{fig:correctionpair}.

\begin{figure} \begin{center}
\includegraphics[width=0.45\textwidth]{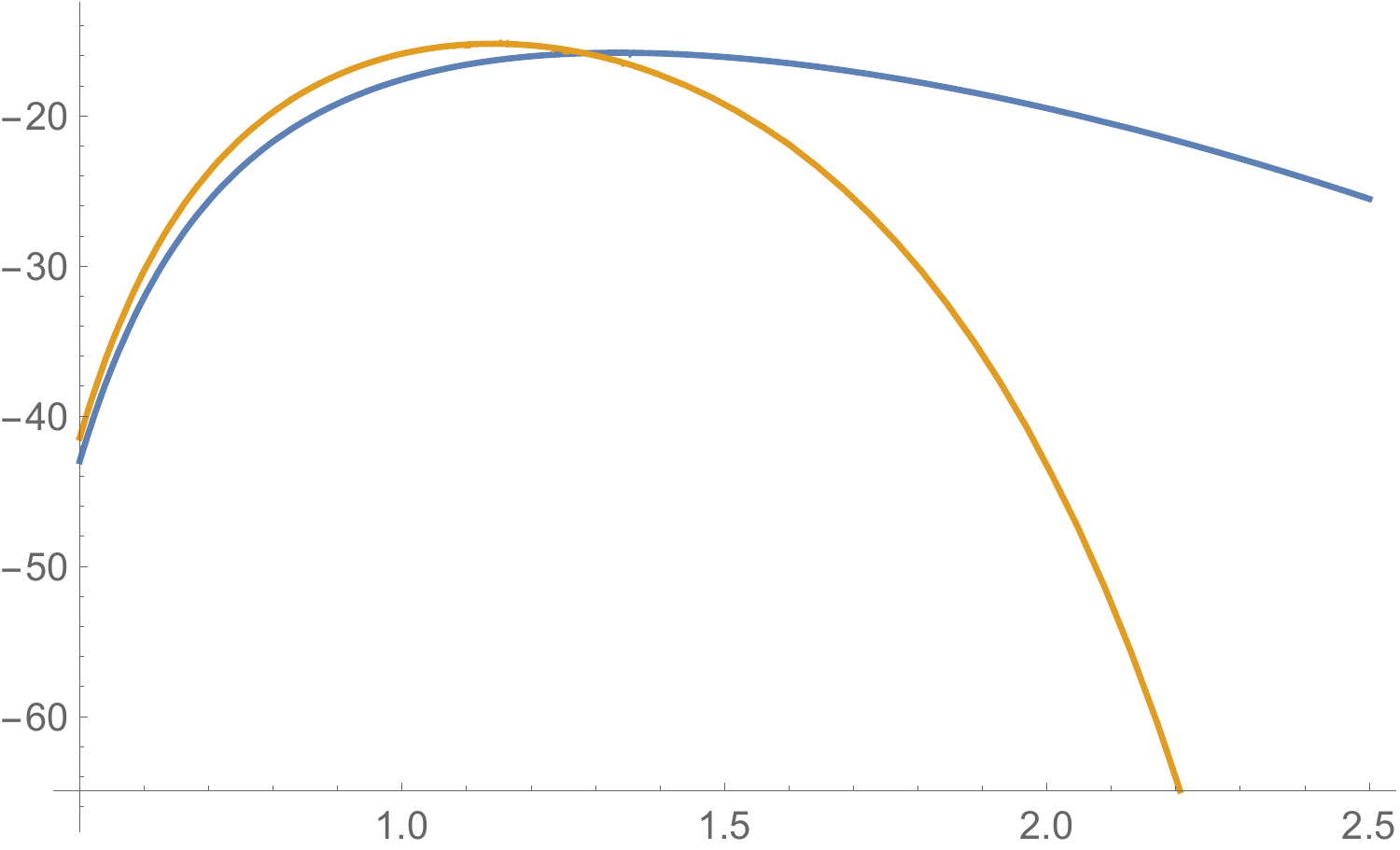} \end{center}
\caption{The integral (\ref{intcorrpair}) as a function of $|k_2|/k_f$ at positive $k_2/k_1$ (blue line) and negative $k_2/k_1$ (yellow line).}
\label{fig:correctionpair}
\end{figure}

Note that in accordance with Eqs. (\ref{occup},\ref{corrpoi3}) $\delta F \propto k_1^{-4/3}$. Therefore the integral over $k_1$ in the expression for the correction to the second moment of vorticity
\begin{equation}
\delta \langle \varpi^2 \rangle =
\int \frac{d^2 k}{(2\pi)^2} \delta F,
\nonumber
\end{equation}
is sitting on the low limit of the interval (\ref{corrpai2}), $k_1\sim \nu k_f^3/\varSigma$. Therefore at calculating $\delta \langle \varpi^2 \rangle$ one cannot use the estimate (\ref{occup}) or the expression (\ref{gausspc}), and one should return to the general expression for the pair correlation function.

\subsection{Correction to the moments}
\label{subsec:corrmom}

Integrating the expression (\ref{corrpai1}) over $\bm k$ and passing to the integration over $\bm p$, one finds
\begin{eqnarray}
\delta \langle \varpi^2 \rangle
=2\int^{\infty}_0 dt \int_{0}^\infty ds \int \frac{d^2 m\, d^2p}{(2\pi)^4}
\nonumber  \\
G(t,-q_1,-q_2+\varSigma s q_1) G(s+t, \bm q)
\nonumber \\
V(\bm m,\bm p) G(s,\bm p)
G(s,\bm m) F(\bm m)
\nonumber \\
\left[\frac{F(\bm q)}{\bm q^2}+\frac{F(\bm p)-F(\bm q)}{{\bm m}^2}\right]
(m_2p_1-m_1 p_2),
\label{secmom1}
\end{eqnarray}
where $\bm q=\bm p+\bm m$. As for the pair correlation function in the region (\ref{corrpai2}), in the integral (\ref{secmom1}) $m_1\ll m_2$ and $p_1 \ll p_2$. Therefore one can use the substitution (\ref{deltaf}).

Performing the substitution (\ref{deltaf}) in Eq. (\ref{secmom1}), we get
\begin{eqnarray}
\delta \langle \varpi^2 \rangle
=\pi\int^{\infty}_0 dt \int_{0}^\infty ds \int \frac{d^2 m\, d^2p}{(2\pi)^4}
\nonumber  \\
G(t,-q_1,-q_2+\varSigma s q_1) G(s+t, \bm q)
G(s,\bm p) G(s,\bm m)
\nonumber \\
F(\bm m) \left[\frac{1}{|m_1|}\delta(m_2-\varSigma s m_1)
-\frac{1}{|p_1|}\delta(p_2-\varSigma s p_1) \right]
\nonumber \\
\left[\frac{F(\bm q)}{\bm q^2}+\frac{F(\bm p)-F(\bm q)}{{\bm m}^2}\right]
(m_2p_1-m_1 p_2)^2,
\label{secmom2}
\end{eqnarray}
The estimates for the variables in the integral (\ref{secmom2}) are
\begin{eqnarray}
s\sim t\sim \frac{1}{\nu k_f^2}, \quad
p_2\sim m_2 \sim k_f,
\nonumber \\
p_1\sim k_1 \sim \frac{\nu k_f^3}{\varSigma}, \quad
F\sim \frac{\epsilon}{\nu k_f^2},
\nonumber
\end{eqnarray}
where we assumed $\alpha \ll \nu k_f^2$. Then
\begin{eqnarray}
\delta \langle \varpi^2 \rangle
\sim \frac{\epsilon^2 k_f^2}{\nu \varSigma^3}
\sim \beta \frac{1}{\varSigma \tau_\star}\langle \varpi^2\rangle_0,
\label{secmom3}
\end{eqnarray}
where $\beta$ is defined by Eq. (\ref{turb4}).

Analogously, one can find the estimate for the non-diagonal component of Reynolds stress tensor
\begin{equation}
\delta\langle v_1 v_2 \rangle
\sim \frac{\epsilon^2 k_f^2}{\varSigma^4}
\sim \beta \frac{\nu k_f^2}{\varSigma}
\langle v_1 v_2 \rangle_0,
\label{secmom4}
\end{equation}
where $\langle v_1 v_2 \rangle_0$ is given by Eq. (\ref{reyn2}). We see in the estimate (\ref{secmom4}) the additional smallness $\nu k_f^2/\varSigma$. It is in accordance with the  energy balance (\ref{energybal}). Indeed, it follows from Eq. (\ref{energybal}) at $\alpha \ll \nu k_f^2$, that the correction $\delta\langle v_1 v_2 \rangle$ to the value $\epsilon/\varSigma$ should contain an extra factor $\nu /\varSigma$, in comparison with the correction $\delta \langle \varpi^2 \rangle$. Then we obtain Eq. (\ref{secmom4}) from Eq. (\ref{secmom3}).

\subsection{Higher order corrections to the pair correlation function}

Each Feynman diagram has an amount of loops. For example, the diagrams depicted in Fig. \ref{fig:paircorr} are characterized by one loop, whereas the diagram depicted in Fig. \ref{fig:paircorr2} contains two loops. The number of loops correspond to the order of the perturbation series. We can estimate the parameter controlling the loop expansion for the pair correlation function in the region (\ref{corrpai2}). Passing from the diagram with $n$ loops to the diagram with $n+1$ loops, we get two additional vertices, one additional $\mathcal G$-line and one additional $\mathcal F$-line. We get also two additional integrations over times (related to the two additional vertices) and an additional integration over a wave vector.

One can choose for the integration the wave vectors $\bm m$ belonging to one of the vertices. Then the wave vectors belonging to the other additional vertex are expressed via the times and the wave vectors $\bm m$. Then the corresponding factor (\ref{turb1}) looks like in Eq. (\ref{turb11}) and can be substituted as in Eq. (\ref{deltaf}). The same logic as for the first correction leads to the conclusion that the factor carrying by the additional loop is
\begin{equation}
\beta \left(\frac{\nu k_f^3}{\varSigma k_1}\right)^{2/3}.
\label{smallparam}
\end{equation}
Thus, just (\ref{smallparam}) is the small parameter justifying the perturbation series.

If we consider corrections to the moments, like $\delta \langle \varpi^2 \rangle$ or $\delta\langle v_1 v_2 \rangle$, then the integrals, determining the corrections, are sitting on the characteristic first components of the wave vectors $\sim \nu k_f^3/\varSigma$, corresponding to the lower limit of the region (\ref{corrpai2}). There the small parameter of the perturbation series is $\beta$. Therefore one can evaluate the $n$-th order corrections to the second moment as
\begin{eqnarray}
\delta_n \langle \varpi^2 \rangle \sim
\frac{\epsilon k_f^2}{\varSigma} \beta^n,
\nonumber \\
\delta_n\langle v_1 v_2 \rangle\sim
\frac{\epsilon \nu k_f^2}{\varSigma^2}
\beta^n.
\label{highmom}
\end{eqnarray}
However, one should be careful. The ``bare'' contributions to the moments (\ref{reyn2},\ref{secondm}) do not obey the logic established for the corrections. Therefore the estimates (\ref{highmom}) do not cover at $n=0$ the expressions (\ref{reyn2},\ref{secondm}), they appear to be much larger than the estimates (\ref{highmom}) for $n=0$.

\section{Conclusion}
\label{sec:conclusion}

We examined statistical properties of the random flow excited by a relatively weak random force on top of a strong static shear flow. We established that the small parameter controlling the perturbation series in the framework of our model is $\beta$ (\ref{turb4}). Note that the parameter is independent of the correlation length of the pumping force. Corrections to the second moment are determined by the wave vectors where the second component is of the order of the wave vector of pumping $k_2\sim k_f$, whereas the first component is much smaller, $k_1\sim \nu k_f^3/\varSigma$. Thus the main contribution to the corrections are supplied by non-isotropic blobs of vorticity, strongly elongated along the shear velocity.

It is worth noting that the contribution of interaction with fluctuations at the pump length, where both components of the wave vectors are estimated as $ k_f$, contains cancellations, leading to effective locality of interaction in ${\bm k}$-space. Such locality justifies the universality of the behavior of correlation functions in the domain (\ref{corrpai2}). Similar cancellations occur in the perturbative approach to three-dimensional turbulence \cite{Kr1,BelLv}.

Our investigation was motivated by examining the structure of the coherent vortices generated as a consequence of the inverse cascade in two-dimensional turbulent flows in finite boxes. One can estimate the parameter $\beta$ for the flat mean polar velocity profile $U=\sqrt{3\epsilon/\alpha}$ \cite{laurie2014universal} where $\epsilon$ is the energy pumped to the fluid per unit mass and $\alpha$ is the bottom friction coefficient. Then the local mean shear rate is $\varSigma \sim \epsilon ^{1/2} \alpha^{-1/2} r^{-1}$, where $r$ is the distance from the vortex center. Thus, the parameter $\beta$ (\ref{turb4}) is
\begin{equation}
\beta \sim \alpha r^2 \nu^{-1}.
\nonumber
\end{equation}
Since the approximation of the local mean shear flow is correct at $k_f r\gg1$ (where $k_f$ is the characteristic wave vector of pumping), the parameter $\beta$ can be small in some region of distances if $\alpha\ll \nu k_f^2$. The condition can be easily achieved in numerical simulations, but it is hardly achieved in laboratory experiments with thin fluid films. Note, however, that our model by itself can be realized experimentally, if besides the small-scale random pumping some strong shear flow is produced in the fluid. In addition, in three-dimensional systems with strong rotation, turbulence is effectively two-dimensioned \cite{Boff14}. As a result, column vortices may occur, which are almost uniform along the axis of rotation \cite{Ogo20}. For such systems, the effective coefficient of friction can be significantly less than the viscous dissipative parameter.

Returning to the coherent vortices, we conclude that it interesting to analyze the case where the parameter $\beta$ is not small. In the case we encounter the situation of strong interaction typical for turbulent systems. The case needs a special investigation which is outside the current work. Note, however, that the energy balance (\ref{energybal}) is an exact relation valid for the strong interaction regime as well. Thus, we expect that the non-diagonal component of Reynolds stress tensor is equal to $\epsilon/\varSigma$ (at large enough shear rate $\varSigma$) even in the strong interaction regime. Therefore the flat velocity profile $U=\sqrt{3\epsilon/\alpha}$ is universal being independent of the character of the interaction.

In our calculations, we used the specific model where the pumping force is short correlation in time. It enables one to relieve the calculations in comparison with the general case of pumping with finite correlation time. However, we hope that the general consequences obtained in the framework of our particular model are universal and are valid for the general case. Say, the energy balance (\ref{energybal}) is independent of the model. Therefore the conclusions concerning the relation between the corrections to the second moment of vorticity and to the non-diagonal component of Reynolds stress tensor (see Subsection \ref{subsec:corrmom}) are universal as well.

\acknowledgments
I.V.Kolokolov is grateful for support by  the Ministry of Science and Higher Education of the Russian Federation.
The work of V.V.Lebedev is performed in the Laboratory “Modern Hydrodynamics” created in frames of Grant 075-15-2022-1099 of the Ministry of Science and Higher Education of the Russian Federation in Landau Institute for Theoretical Physics of RAS and is supported by Grant 23-72-30006 of Russian Science Foundadtion.

\end{document}